\documentclass[twocolumn,superscriptaddress,floatfix,longbibliography,aps,prb,preprintnumbers,10pt,groupedaddress]{revtex4-2}
\usepackage{titlesec}

\usepackage{graphicx}
\usepackage{amssymb}
\usepackage{bm}
\usepackage{graphicx,color}

\usepackage[urlcolor=blue,colorlinks=true,citecolor=blue,linkcolor=blue,pdfstartview={FitH},bookmarks=false]{hyperref}
\begin{document}

\title{Reply to the comment on ``Instability of the ferromagnetic quantum critical point and symmetry of the ferromagnetic ground state in 2D and 3D electron gases with arbitrary spin-orbit splitting''} 

\date{\today}
\author{Dmitry Miserev,$^{1\ast}$ Daniel Loss,$^{1}$ and Jelena Klinovaja$^{1}$}
\affiliation{$^{1}$Department of Physics, University of Basel, \\
	Klingelbergstrasse 82, CH-4056 Basel, Switzerland\\
}

\begin{abstract}
	We reply on the comment of D. Belitz and T. R. Kirkpatrick on our article ``Instability of the ferromagnetic quantum critical point and symmetry of the ferromagnetic ground state in 2D and 3D electron gases with arbitrary spin-orbit splitting''. They correctly pointed out that non-analyticities in the thermodynamic potential of 2D and 3D Fermi liquids that originate from the backscattering processes are suppressed by the square of the Cooper logarithm. This suppression is not so important in 2D Fermi liquids where the non-analyticity is cubic and, therefore, is much more relevant than the quartic Ginzburg-Landau term. However, the Cooper renormalization in 3D Fermi liquids renders marginally relevant quartic non-analyticity to marginally irrelevant one. We argue that the leading $\Omega \propto B^4 \ln(E_F/B)$ form of the 3D non-analyticity in the thermodynamic potential $\Omega$ as a function of Zeeman field $B$ survives due to contribution coming from the arbitrary-angle scattering that is not subject to the Cooper renormalization. In perturbation theory, such correction first emerges in third order in interaction, as was identified by Maslov and Chubukov [Phys. Rev. B \textbf{79}, 075112 (2009)]. However, the sign of this third-order correction is positive, which contradicts to the physical argument on the sign of non-analytic corrections presented by D. Belitz and T. R. Kirkpatrick. Maslov and Chubukov argued that overall sign of the non-analytic correction is inconclusive, and strongly depends on the angular structure of the Landau scattering amplitude in the regime when interaction is no longer weak, this is also the regime where ferromagnetic quantum phase transition is expected. They also showed that the non-analyticity remains marginally relevant in the vicinity of 3D ferromagnetic quantum critical point, $\Omega \propto -B^4 \ln\ln(E_F/B)$. Though this non-analyticity can formally destabilize the $T = 0$ ferromagnetic quantum critical point, it becomes irrelevant at any experimentally achievable temperatures.
\end{abstract}

\maketitle

We  appreciate the comment of D. Belitz and T. R. Kirkpatrick on our article ``Instability of the ferromagnetic quantum critical point and symmetry of the ferromagnetic ground state in 2D and 3D electron gases with arbitrary spin-orbit splitting'' \cite{miserev_instability_2022}. 
Our main result that is represented by our Eqs.~(38), (44), (48), corresponds to the non-analytic correction to the thermodynamic potential $\Omega$ calculated for a short-range interaction with the matrix element $u$ within second order perturbation theory in $u$ for the case of a Fermi surface that is split by an arbitrary spin-orbit interaction and a finite Zeeman field. 
The novelty of our result is in the explicit form of the angular-dependent form-factors enveloping the non-analytic contributions.

The drawback of the perturbative approach we used is missing renormalizations of the Cooper amplitude due to higher order terms in $u$.
At zero Zeeman field $B = 0$ and finite spin-orbit interaction, there is one remaining zero-momentum Cooper channel that respects time reversal symmetry and therefore is unaffected by the spin-orbit splitting of the Fermi surface: the corresponding interaction matrix element is given by the scattering of a spin-singlet pair of electrons with momenta $\bm k_+, -\bm k_+$ into the spin-singlet pair with momenta $-\bm k_-, \bm k_-$, where $k_\pm = |\bm k_\pm|$ are the Fermi momenta of the spin-split Fermi surface, and $\bm k_+$ is parallel to $\bm k_-$.
Other backscattering channels are suppressed as they are set off  resonance due to the momentum mismatch $\propto \beta_{\mathrm{SO}}$ between the scattered pairs, where $\beta_{\mathrm{SO}}$ is the spin-orbit splitting (in order to calculate the contribution of these channels to the spin susceptibility at $B = 0$, an infinitesimal Zeeman field has to be included in calculations of $\Omega$ to enable spin-flip transitions).
A small but finite Zeeman field $B$ breaks time reversal symmetry by boosting a finite momentum mismatch $\propto B$ between the spin-singlet pairs, a mechanism similar to the one that destroys conventional superconductivity. 
However, in the limit $B \to 0$ the logarithmic Cooper renormalizations of the time-reversal-invariant amplitude, $u_{\mathrm{C}}$, at zero temperature $T = 0$, is suppressed by a large logarithm, 
\begin{eqnarray}
	&& u_{\mathrm{C}} = \frac{u}{1 + u \ln(E_F/B)} \to \frac{1}{\ln(E_F/B)} \, , \label{u}
\end{eqnarray}
where $E_F$ is the Fermi energy, $B$ the absolute value of the Zeeman field, $u > 0$ is the bare matrix element of the backscattering interaction in the zero-momentum spin-singlet channel.  
This renormalization is precisely what was pointed out by D. Belitz and T. R. Kirkpatrick in their comment, and we fully agree with them on this matter.

The effects of such Cooper renormalizations given by Eq.~(\ref{u}) are well-known and have been broadly studied in the context of non-analytic corrections to $\Omega$ for 2D and 3D Fermi liquids. 
For example, extensive non-perturbative treatment of corrections non-analytic in temperature $T$  to the heat capacity of interacting spin-degenerate Fermi liquids results in a very similar non-analyticity of $\Omega$ as a function of $T$ yielding the non-analytic corrections to the heat capacity, see Eq.~(7.37a) for 2D, and Eq.~(7.37b) for 3D Fermi liquids in Ref.~\cite{aleiner_supersymmetric_2006}.
Corrections originating from the Cooper channel are suppressed by the logarithm squared [denoted by $\mathcal{X}(T)$] compared to the perturbative result (see Eqs.~(7.29a), (7.29b) in~\cite{aleiner_supersymmetric_2006}).
The $T$ dependence of the spin susceptibility at $B = 0$ originates from the spin-singlet Cooper channel exclusively, and is given by Es.~(7.42) for 2D, and Eq.~(7.43) for 3D Fermi liquids in Ref.~\cite{schwiete_temperature_2006} in the limit $T \to 0$, yet again demonstrating $\propto 1/\ln^2(E_F/T)$ suppression compared to the perturbative result given by Eqs.~(7.40), (7.41). 
The $B$-dependence of the spin susceptibility at $T = 0$ originating from the backscattering processes suffers similar suppression by $1/\ln^2(E_F/B)$.

However, such logarithmic suppressions are not essential for studying two-dimensional (2D) quantum phase transitions as the $B^3$ non-analyticity, even scaled down by $\propto 1/\ln^2(E_F/B)$, still represents the leading non-analyticity that is more relevant than the analytic $\propto B^4$ term in $\Omega(B)$.
In particular, Eq.~(44) in our work \cite{miserev_instability_2022} still represents the relevant non-analyticity in the 2D thermodynamic potential, though each term in Eq.~(44) should be scaled down by the square of the Cooper logarithm with the appropriate argument.
In particular, the $\propto B^3$ non-analyticity is scaled down to $\propto B^3/\ln^2(E_F/\mathrm{max}[B, T])$ at exponentially small $B$ and $T$ satisfying $u \ln(E_F/\mathrm{max}[B, T]) \gg 1$.
However, such exponentially small values of $B$ and $T$ might not even be relevant for studying 2D ferromagnetic quantum phase transitions unless the scale $B_0$ at which the non-analytic correction in $\Omega$ is balanced by the Ginzburg-Landau $\propto B^4$ term is also exponentially small.
For this reason, the Cooper renormalization of the cubic non-analyticity cannot stabilize the 2D ferromagnetic quantum critical point (QCP).

The Cooper renormalization is much more serious for 3D metals, where the leading non-analyticity in $\Omega$ scales down from the marginally relevant $\propto B^4 \ln (E_F/B)$ (perturbative result) to marginally irrelevant $\propto B^4/\ln (E_F/B)$ (renormalized non-analyticity) which qualitatively changes the role of such non-analyticity: indeed, at $B \to 0$ the $\propto B^4/\ln (E_F/B)$ non-analyticity is negligible compared to the Ginzburg-Landau $\propto B^4$ term.

D. L. Maslov and A. V. Chubukov explicitly demonstrated in Ref.~\cite{maslov_nonanalytic_2009} that higher-order corrections to the nonanalyticity in the thermodynamic potential can be separated into two classes: corrections originating from the backscattering terms (these are the ones that are suppressed by the Cooper renormalization), and corrections originating from scattering at arbitrary angle (they refer to these processes as uncorrelated momentum scattering). 
The backscattering terms can be treated non-perturbatively in a systematic way using the diagrammatic resummation as done in Ref.~\cite{maslov_nonanalytic_2009} or via the multidimensional bosonization procedure developed by I. L. Aleiner, G. Schwiete, and K. B. Efetov in Refs.~\cite{aleiner_supersymmetric_2006,schwiete_temperature_2006}. 
Importantly, the same $\propto B^4 \ln (E_F/B)$ non-analyticity in 3D $\Omega$  emerges also from scattering at arbitrary angle.
The first non-trivial correction that does not belong to the backscattering class is shown in Fig.~2e in Ref.~\cite{maslov_nonanalytic_2009} yielding  negative $\delta \chi \propto -B^2 \ln(E_F/B)$ correction to the spin susceptibility given by Eq.~(5.8).
Though this correction is not subject to the Cooper renormalization, its negative sign renders it irrelevant at small interaction $u \ll 1$.
However, in the vicinity of the ferromagnetic QCP the interaction coupling constant is no longer small, $u_{\mathrm{QCP}} \sim 1$, which rises the question of the Fermi-liquid renormalizations of such non-backscattering contributions at finite interaction strength $u$.
D. L. Maslov and A. V. Chubukov  shed  light on these renormalizations in Ref.~\cite{maslov_nonanalytic_2009} by using the two-component Landau scattering amplitude given by Eq.~(2.39).
They demonstrated that even in case of weakly anisotropic scattering amplitude, the sign of the 2D non-analytic correction can be reversed, see Eq.~(2.51) and Fig.~4 for the prefactor of the 2D spin susceptibility.
They also argue that the sign of the 3D non-analytic spin susceptibility is non-universal and depends on the form of the Landau scattering amplitude.
Unfortunately, the full calculation of the prefactor for the 3D non-analyticity in the relevant regime of strongly anisotropic scattering amplitude that is large for the forward scattering and vanishes for the backscattering, is challenging and remains an open problem.
Instead, they calculated the non-analytic 3D spin susceptibility in the vicinity of the Pomeranchuk ferromagnetic QCP, and found an extremely weak non-analyticity, $\delta \chi \propto + B^2 \ln\ln(E_F/B)$, which is formally relevant at $T = 0$, yet practically irrelevant at any realistic temperatures, see Eq.~(5.14) in Ref.~\cite{maslov_nonanalytic_2009}.

In conclusion, the 3D non-analyticity of the thermodynamic potential, $\Omega \propto B^4 \ln(E_F/B)$, originates not only from the backscattering contribution that is subject to Cooper renormalizations, but also from the arbitrary-angle scattering processes that retain 
the 3D non-analyticity $\Omega \propto B^4 \ln(E_F/B)$ without Cooper renormalization.
The sign of such 3D non-analyticities is not conclusive and depends on the angular structure of the Landau scattering amplitude, according to calculations performed in Ref.~\cite{maslov_nonanalytic_2009}.
In the vicinity of a 3D ferromagnetic QCP, the non-analyticity in $\Omega$ weakens to $\propto -B^4 \ln\ln(E_F/B)$ yet remains marginally relevant at $T = 0$.
As the $\Omega \propto B^4 \ln(E_F/B)$ non-analyticity originating from the arbitrary-angle scattering is not subject to the Cooper renormalization, and therefore dominates over
the Ginzburg-Landau  term $\propto B^4$, the physical argument in favor of negative non-analyticity in the 3D spin susceptibility used by D. Belitz and T. R. Kirkpatrick is not at all clear in this context.
The status of the sign of this non-analytic correction at realistic interaction strength $u \sim 1$ at which the ferromagnetic QCP is expected, remains an open question.
This was the main reason why we concentrated on the second-order correction and its anisotropy induced by the interplay between the spin-orbit coupling and the Zeeman field.
At the same time, the 2D non-analyticity $\Omega \propto B^3$ is much stronger compared to the analytic $\propto B^4$ term, and provides a second minimum at magnetization $B_0$ that is not exponentially small. 
For this reason, the Cooper renormalization of 2D non-analyticity at relevant values of $B \sim B_0$ is negligible.

%

\end{document}